\documentclass[aps,prl,twocolumn]{revtex4}

\usepackage{graphicx}% Include figure files
\usepackage{bm}% bold math
\usepackage{color}

\newcommand{\be}{\begin{equation}}
\newcommand{\ee}{\end{equation}}
\newcommand{\bea}{\vspace{0.25cm}\begin{eqnarray}}
\newcommand{\eea}{\end{eqnarray}}

\begin{document}
\title{Measuring incompatible observables of a single photon}

\author{F.~Piacentini$^{1}$, M.~P.~Levi$^{1,2}$, A.~Avella$^{1}$, E.~Cohen$^{3}$,  R.~Lussana$^{4}$, F.~Villa$^{4}$, A.~Tosi$^{4}$, F.~Zappa$^{4}$, M.~Gramegna$^{1}$, G.~Brida$^{1}$, I.~P.~Degiovanni$^{1}$, M.~Genovese$^{1,5}$}

\affiliation{$^{1}$INRIM, Strada delle Cacce 91, I-10135 Torino, Italy}
\affiliation{$^{2}$Politecnico di Torino, Corso Duca degli Abruzzi 24, I-10129Torino, Italy}
\affiliation{$^{3}$School of Physics and Astronomy, Tel Aviv University, Tel Aviv 6997801, Israel}
\affiliation{$^{4}$Politecnico di Milano, Dipartimento di Elettronica, Informazione e Bioingegneria, Piazza Leonardo da Vinci 32, 20133 Milano, Italy}
\affiliation{$^{5}$INFN, Via P. Giuria 1, I-10125 Torino, Italy}

\begin{abstract}
One of the most intriguing aspects of Quantum Mechanics is the impossibility of measuring at the same time observables corresponding to non-commuting operators.
This impossibility can be partially relaxed when considering joint or sequential weak values evaluation \cite{2,1,1b,steinbergPRL,seq}. Indeed, weak measurements have been a real breakthrough in the quantum measurement framework that is of the utmost interest from both a fundamental \cite{4,4a,4b,4c,3, 3a, 4d,4e} and  an applicative point of view \cite{4f,5,6,6a,6b,6c,7a,7b}.
In this paper, we show how we realized for the first time a sequential weak value evaluation of two incompatible observables on a single photon.
\end{abstract}
\maketitle

%\section{Introduction}
Measurements are the very basis of Physics. In Quantum Mechanics they assume even a more fundamental role, since observables can have undetermined values that ``collapse'' on a specific one only when a strong measurement (described by a projection operator) is performed. Furthermore, a crucial feature of quantum measurement is that measuring one observable completely erases the information on the corresponding conjugate one (e.g. measurement of position erases information about momentum).
Weak values, introduced in \cite{2} and firstly realized in \cite{3,3a,5}, represent a new quantum measurement paradigm, where only a small amount of information is extracted from a single measurement, so that the state basically does not collapse.
They can have anomalous values (imaginary, unbounded values) and, while their real part is usually interpreted as a conditioned average of the observable in the limit of zero disturbance \cite{w1}, their imaginary part is related to the disturbance (or backaction) of the measuring pointer during the measurement process \cite{w2}. Furthermore, every Positive Operator Valued Measure (POVM) can be realised as a sequence of weak measurements \cite{w3}.
Weak measurements have been used for addressing fundamental questions \cite{4} such as contextuality \cite{4e}, but can also be seen as a groundbreaking tool for quantum metrology allowing high-precision measurements (at least in presence of specific noises \cite{4f}), as the tiny spin Hall effect \cite{5} or small beam deflections \cite{6} and characterization of quantum states \cite{7a,7b}.
One of the most intriguing properties of weak measurements is that, since they do not make the wave function collapse, they permit measuring simultaneously non-commuting observables, challenging ``one of the canonical dicta of quantum mechanics" \cite{seq}.

Here we demonstrate for the first time this possibility by evaluating a sequential weak value of non-commuting observables of heralded single photon polarisation states.

Specifically, the weak value of an observable $\widehat{A}$ is defined as $ \langle \widehat{A} \rangle _w = { \langle \psi_f | \widehat{A} |\psi_i \rangle \over \langle \psi_f | \psi_i \rangle}$, where the key role is symmetrically played by the pre-selected ($ | \psi_i \rangle$) and post-selected ($|\psi_f \rangle$) quantum states. When the pre- and post-selected states are equal, the weak value is just the expectation value of $\widehat{A}$.

When a von Neumann coupling is considered between the observable $\widehat{A}$ and a pointer observable $\widehat{P}$, according to the unitary transformation $\widehat{U}= \exp (- i g \widehat{A} \otimes \widehat{P})$, and the weak interaction regime is assumed, one can describe the evolution of this system, prepared in the pre-selected state and projected on the post-selected state, as
\begin{equation}
\langle \psi_f | e^{- i g \widehat{A} \otimes \widehat{P}}  |\psi_i \rangle \simeq  \langle \psi_f  |\psi_i \rangle (\mathbf{1} -i g \langle \widehat{A} \rangle_w \widehat{P}) .
\end{equation}
In our case, a polarisation projector is the observable coupled to the pointer variable represented by the momentum operator $\widehat{P}$, which is the generator of displacement of the single-photon transverse spatial wave function. This spatial displacement - due to the polarisation-sensitive spatial walk-off of the Poynting vector of the single photon induced by its propagation into a birefringent medium - realises in practice the weak interaction. Thus, by measuring the position observable $\widehat{X}$, canonically conjugated to the pointer observable $\widehat{P}$, after the pre- and post-selection of the single-photon polarisation state, one obtains $\langle \widehat{X} \rangle =g  \langle \widehat{A} \rangle _w  $.

Measurements of joint \cite{steinbergPRL} or sequential \cite{seq}  weak values of two observable $\widehat{A}$ and $\widehat{B}$ are obtained when two different couplings ($g_x$ and $g_y$) to two distinct pointer observables (in our experiment the two transverse momenta $\widehat{P}_x$ and $\widehat{P}_y$ ) are realised between the pre- and post-selection of the state. In particular, if the measurement is performed exploiting simultaneous interactions, we are dealing with measurement of the joint weak value, and by measuring the covariance of the position observables $\widehat{X}$ and $\widehat{Y}$ ($\langle \widehat{X} \widehat{Y} \rangle $) one obtains \cite{steinbergPRL}
\be
\langle \widehat{X} \widehat{Y} \rangle = \frac{1}{4} g_x g_y \mathrm{Re} \left[  \langle \widehat{A} \widehat{B} +\widehat{A} \widehat{B}  \rangle_w + 2 \langle \widehat{A}  \rangle_{w}^{*}  \langle \widehat{B}  \rangle_{w}  \right],  \label{joint1}
\ee
while if we have a sequence of two weak interactions, e.g. the first interaction is described by the unitary transformation $\widehat{U}_x= \exp( - i g_x \widehat{A} \otimes \widehat{P}_x )$ and the second by $\widehat{U}_y= \exp (- i g_y \widehat{B} \otimes \widehat{P}_y)$, when measuring  $\langle \widehat{X} \widehat{Y} \rangle $ one obtains \cite{seq}
\be
\langle \widehat{X} \widehat{Y} \rangle = \frac{1}{2} g_x g_y \mathrm{Re} \left[  \langle \widehat{A} \widehat{B}   \rangle_w +  \langle \widehat{A}  \rangle_{w}^{*}  \langle \widehat{B}  \rangle_{w}  \right].  \label{sequential1}
\ee

Thus, the real part of sequential ($\mathrm{Re}[  \langle \widehat{A} \widehat{B}   \rangle_w ]$ ) or joint ($\mathrm{Re}[  \langle \widehat{A} \widehat{B}+ \widehat{B} \widehat{A}   \rangle_w]$) weak values can be evaluated by the measurement of $\langle \widehat{X} \widehat{Y} \rangle$ and by the evaluation of each weak value independently, i.e. of $\langle \widehat{A}   \rangle_w $  and  $\langle \widehat{B}   \rangle_w $ (these can be obtained by measuring the mean values of the positions and momenta independently, namely $\langle \widehat{X} \rangle$, $\langle \widehat{Y} \rangle$, $\langle \widehat{P}_x  \rangle$ and $\langle \widehat{P}_y  \rangle$ \cite{steinbergPRL, seq}).

\begin{figure}[tbp]
\begin{center}
\includegraphics[width=0.45\textwidth]{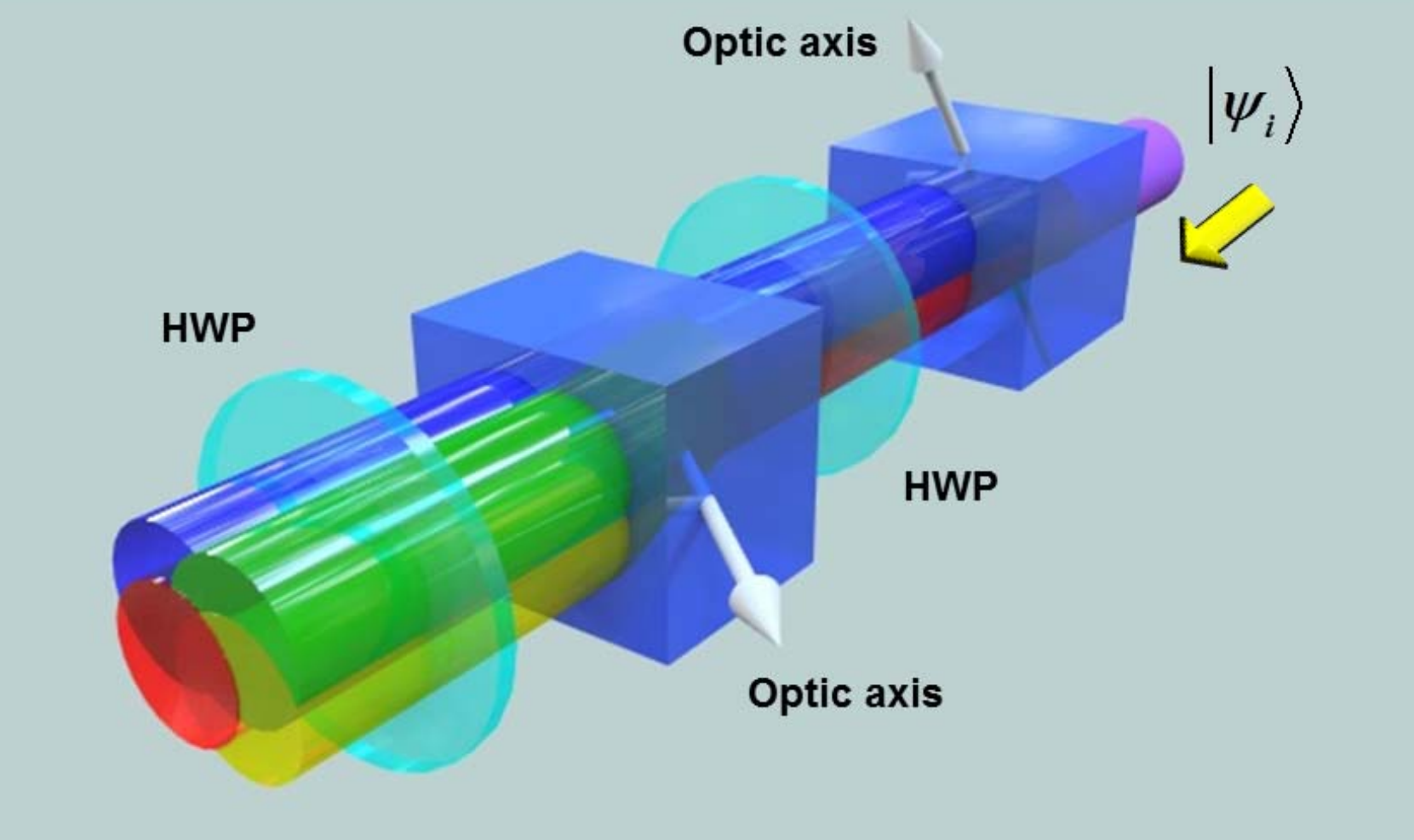}
\caption{Sequential weak interactions performed on the single-photon polarisation state $|\psi_i\rangle$, by means of two birefringent crystals with orthogonal optical planes. HWP: Half Wave Plate.
} \label{joint_weak3D}
\end{center}
\end{figure}

In our experiment we focus on the case of sequential weak values evaluation (see Fig. \ref{joint_weak3D}), where the meter operators $\widehat{A}$ and $\widehat{B}$ are the linear projectors $\widehat{\Pi}_V = | V \rangle \langle V |$ and $\widehat{\Pi}_\psi = | \psi \rangle \langle \psi |$ (with $|\psi \rangle = \cos \theta |H \rangle + \sin \theta |V \rangle $).

To perform our experiment we exploited a heralded single-photon source based on pulsed parametric down-conversion (see Fig. \ref{setup} and Setup section). The heralded single-photon is prepared (pre-selected) in the linear polarisation state $|\psi_ {i}\rangle$, then it undergoes the two sequential weak interactions in two birefringent media having perpendicular optical planes.
The two weak interactions induce walk-off effects on the transverse spatial degree of freedom of the single photon state according to the two unitary transformations $\widehat{U}_y= \exp (-i g_y \widehat{\Pi}_V \otimes \widehat{P}_y)$, and $\widehat{U}_x= \exp (-i g_x \widehat{\Pi}_\psi \otimes \widehat{P}_x)$, (see Fig. \ref{joint_weak3D} and Theory section for details).

To evaluate the sequential weak value of the (in general) non-commuting projectors $\langle \widehat{\Pi}_\psi \widehat{\Pi}_V \rangle_w  $, as well as the single meter weak values $\langle \widehat{\Pi}_\psi \rangle_w  $ and $\langle  \widehat{\Pi}_V \rangle_w  $, it is enough to measure the mean transverse position observables $\widehat{X}$ and $\widehat{Y}$ (conjugate of the pointer momentum observables $\widehat{P}_x $ and $\widehat{P}_y $, respectively). This is possible in our experiment since we are dealing with linear polarisations only.  In fact, according to Eq. (\ref{sequential1}) we have $\langle \widehat{X} \widehat{Y} \rangle = \frac{1}{2} g_x g_y  \left(  \langle \widehat{\Pi}_{\psi} \widehat{\Pi}_V   \rangle_w +  \langle \widehat{\Pi}_{\psi}  \rangle_{w}  \langle \widehat{\Pi}_V  \rangle_{w}  \right)$, $\langle \widehat{X}  \rangle = g_x \langle \widehat{\Pi}_{\psi}  \rangle_{w}  $, $\langle \widehat{Y}  \rangle = g_y \langle \widehat{\Pi}_{V}  \rangle_{w}  $.

The main results of our work are summarised in Fig. \ref{results}.
Here we plot the two weak values and the joint sequential weak measurement as a function of the polarization selection of the second measurement (the first one being kept fixed), compared with the theoretical prediction.

Our data clearly demonstrate the possibility of obtaining a sequential weak measurement of incompatible observables, both for anomalous and non-anomalous weak values, proving a good agreement with theory. This is, as described in \cite{seq}, a breakthrough in quantum measurement experiments paving the way to new protocols in quantum sensing and quantum metrology.

In the end, it is worth mentioning that this experiment does not only shed light on controversial issues like ``the past of quantum systems'' \cite{past} and counterfactual computation \cite{counter1, counter2},  but in fact enables for the first time their experimental test (which was indeed the original aim of the theoretical work in \cite{seq}). This also holds for the several possible applications that have arisen since then, including quantum process tomography \cite{QPT}.
%, quantum correlation functions \cite{QCF}.
% and tests of recent uncertainty relations \cite{WUR}.

\section{Theory }

Our single photon state is prepared (pre-selected) in the initial state $|\phi_i \rangle \rangle= |\psi_i\rangle \otimes |f_x \rangle \otimes |f_y \rangle  $, with $ |\psi_i\rangle = \cos \theta_i | H \rangle +\sin \theta_i | V \rangle  $ and $ |f_\xi \rangle = \int \mathrm{d} \zeta \mathcal{F}_\xi (\zeta) |\zeta \rangle$, where $|\mathcal{F}_\xi (\zeta)|^2  $ is the probability density function of detecting the photon in the position $\xi$ (with $\xi=x,y$) of the transverse spatial plane. $|\mathcal{F}_\xi (\zeta)|^2  $ in our experiment is reasonably Gaussian, since the single photon guided in a single-mode optical fiber is collimated with a telescopic optical system. By experimental evidence, we can assume that the (unperturbed) $|\mathcal{F}_\xi (\zeta)|^2  $ is centered around zero and has the same width $\sigma$ both for $\xi=x$ and for $\xi=y.$

The single photon undergoes the weak interactions that manifest themselves as a spatial walk-off induced by the birefringence in optical crystals. This sequence of weak interactions is described by the unitary transformations $\widehat{U}_y= \exp (-i g_y \widehat{\Pi}_V \otimes \widehat{P}_y)$ and $\widehat{U}_x= \exp (-i g_x \widehat{\Pi}_\psi \otimes \widehat{P}_x)$. Then the single-photon is projected on the post-selected state $|\psi_f \rangle$ and detected by a spatial-resolving detector. The post-selected single-photon state observed by the spatial-resolving detector is $|\phi_f \rangle \rangle = \langle \psi_f| \widehat{U}_x \widehat{U}_y | \psi_i \rangle \rangle$. Thus, the mean values of the positions of the single-photon detected after the post-selection are $\langle \widehat{X} \rangle_f = g_x \langle \widehat{\Pi}_\psi \rangle_w $ and $\langle \widehat{Y} \rangle_f = g_y \langle \widehat{\Pi}_V \rangle_w $. The expected value of the covariance of the $X$ and $Y$ positions of the single photon detected after the post-selection is $\langle \widehat{X} \widehat{Y} \rangle_f = \frac{g_x g_y }{2} (\langle \widehat{\Pi}_\psi \widehat{\Pi}_V \rangle_w + \langle \widehat{\Pi}_\psi  \rangle_w   \langle  \widehat{\Pi}_V \rangle_w) $. By inverting these simple relations it is possible to obtain the weak values of the two incompatible observables $\langle \widehat{\Pi}_V \rangle_w$ and $\langle \widehat{\Pi}_\psi \rangle_w$,  as well as the sequential weak value of the two incompatible observables $\langle \widehat{\Pi}_\psi \widehat{\Pi}_V \rangle_w$. Note that this relation between position mean values and polarisation weak values holds only in the case of weak interaction, i.e. only for $g/\sigma \ll 1$. In our case we have evaluated $g_x/\sigma \sim g_y/\sigma \sim 0.15 $

\section{Setup}

Our experimental setup (Fig. \ref{setup}) is constituted of a 796~nm mode-locked Ti:Sapphire laser (repetition rate: 76 MHz), whose second harmonic emission pumps a $10\times10\times5$ mm LiIO$_3$ nonlinear crystal, producing Type-I Parametric Down-Conversion (PDC).

The idler photon ($\lambda_i=920$ nm) is coupled to a single-mode fiber (SMF) and then addressed to a Silicon Single-Photon Avalanche Detector (SPAD), heralding the presence of the correlated signal photon ($\lambda_s=702$ nm) that, after being SMF-coupled, is sent to a launcher and then to the free-space optical path, where the experiment for weak values evaluation is performed.

We have estimated the quality of our single-photon emission obtaining a $g^{(2)}$ value (or more properly a parameter $\alpha$ value \cite{grangier}) of ($0.13\pm0.01$) without any background/dark-count subtraction.

After the launcher, the heralded single photon state is collimated by a telescopic system, and then prepared (pre-selected) in a linear polarization state $|\psi_{i}\rangle$ (by means of a calcite polarizer followed by a half-wave plate). The first weak interaction is carried out by a 1 mm long birefringent crystal (BC$_V$) whose extraordinary ($e$) optical axis lies in the $Y$-$Z$ plane, with an angle of $\pi/4$ with respect to the $Z$ direction. Due to the spatial walk-off effect experienced by the vertically-polarized photons  (i.e. along $Y$ direction), horizontal and vertical-polarization paths get slightly separated along the $Y$ direction, inducing in the initial state $|\psi_{i}\rangle$ a small decoherence (below $5\%$) that keeps it substantially unaffected.

After a phase compensation tuned in order to nullify the temporal walk-off in BC$_V$, the photon goes to the second weak measurement system. It is constituted by a 1 mm long birefringent crystal with the optical $e$-axis lying in the $X$-$Z$ plane (BC$_H$), inserted between two half-wave plates. By rotating both wave-plates of the same angle with respect to the $H$-axis, one obtains the weak interaction on the linear polarisation state $| \psi \rangle$ with the polarisations separation appearing along the $X$ direction.

After both weak measurements are performed (each with its corresponding phase compensation), the photon meets a half-wave plate and a calcite polarizer, used to project the state onto the post-selected state $|\psi_{f}\rangle$, and then it is detected by a spatial-resolving single-photon detector prototype. This device is a two-dimensional array made of 32x32 ``smart pixels'' -each pixel includes a SPAD detector and its front-end electronics for counting and timing single photons \cite{VILLA2014}-. All the pixels operate in parallel with a global shutter readout.
The SPAD array is gated with 6 ns integration windows, triggered by the SPAD detector of the heralding arm. Therefore, since the heralding detection rate is in the order of $100$ kHz, the effective dark count rate of the array is drastically reduced by the low duty cycle, thus improving the signal-to-noise ratio.

{\bf Acknowledgements}

This work has been supported by EMPIR-14IND05 ``MIQC2'' (the EMPIR initiative is co-funded by the EU H2020 and the EMPIR Participating States) and the EU FP7 project ``BRISQ2''. E.C. was supported in part by the Israel Science Foundation Grant No. 1311/14. We wish to thank Yakir Aharonov and Avshalom C. Elitzur for helpful discussions.

\begin{widetext} $ $
\begin{figure}[tbp]
\begin{center}
\includegraphics[width=0.95\textwidth]{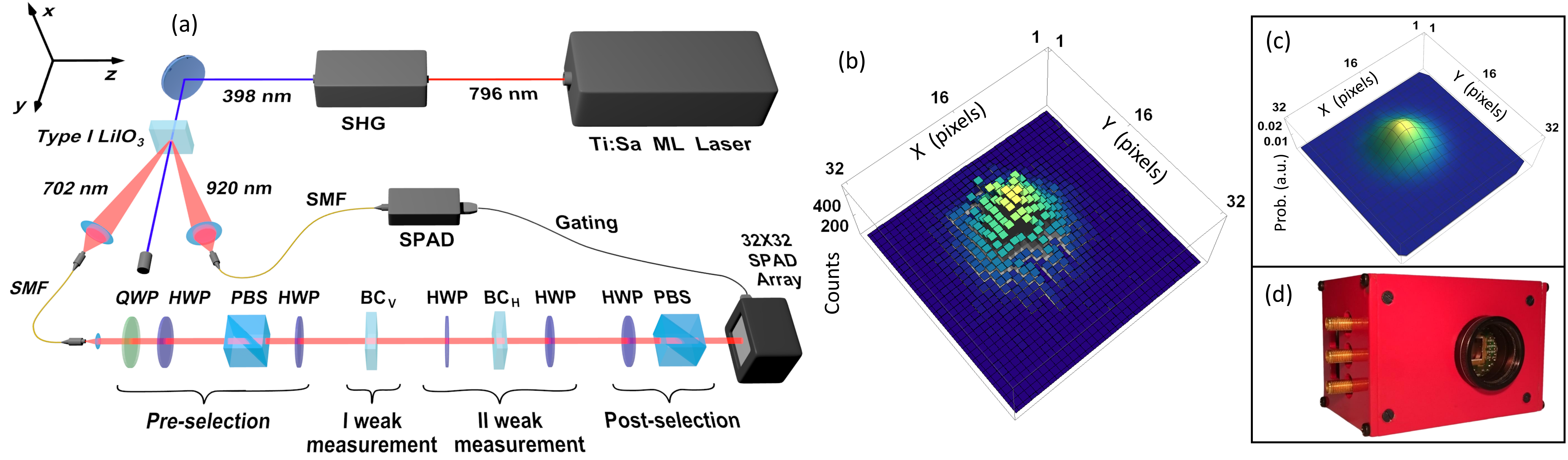}
\caption{(a) A frequency doubled mode-locked Ti:Sa laser pumps a LiIO$_3$ Type-I PDC crystal. The idler photon ($\lambda_i=920$ nm) is coupled to a single-mode fiber (SMF) and addressed to a SPAD heralding the correlated signal photon ($\lambda_s=702$ nm) that is prepared in a linear polarization state (pre-selection block) and sent to the in-air weak measurement apparatus.
The first weak measurement is operated by the birefringent crystal BC$_V$, followed by the BC$_H$ block, in which the second weak measurement takes place. Finally, the photon is post-selected and detected (SHG: Second Harmonic Generator; QWP: Quarter Wave Plate; HWP: Half Wave Plate; PBS: Polarizing Beam Splitter; BC: Birefringent Crystal).
(b) Typical single data acquisition obtained with our spatial resolving single-photon detector (32X32 SPAD camera), after noise subtraction. It represents the number of counts acquired in 300 s versus the different pixels of the SPAD array.
(c) The corresponding predicted probability distribution calculated according to the theory.
(d) SPAD camera prototype used in the experiment.
} \label{setup} \end{center} \end{figure}
\end{widetext}

\begin{widetext} $ $
\begin{figure}[tbp]
\begin{center}
\includegraphics[width=0.9\textwidth]{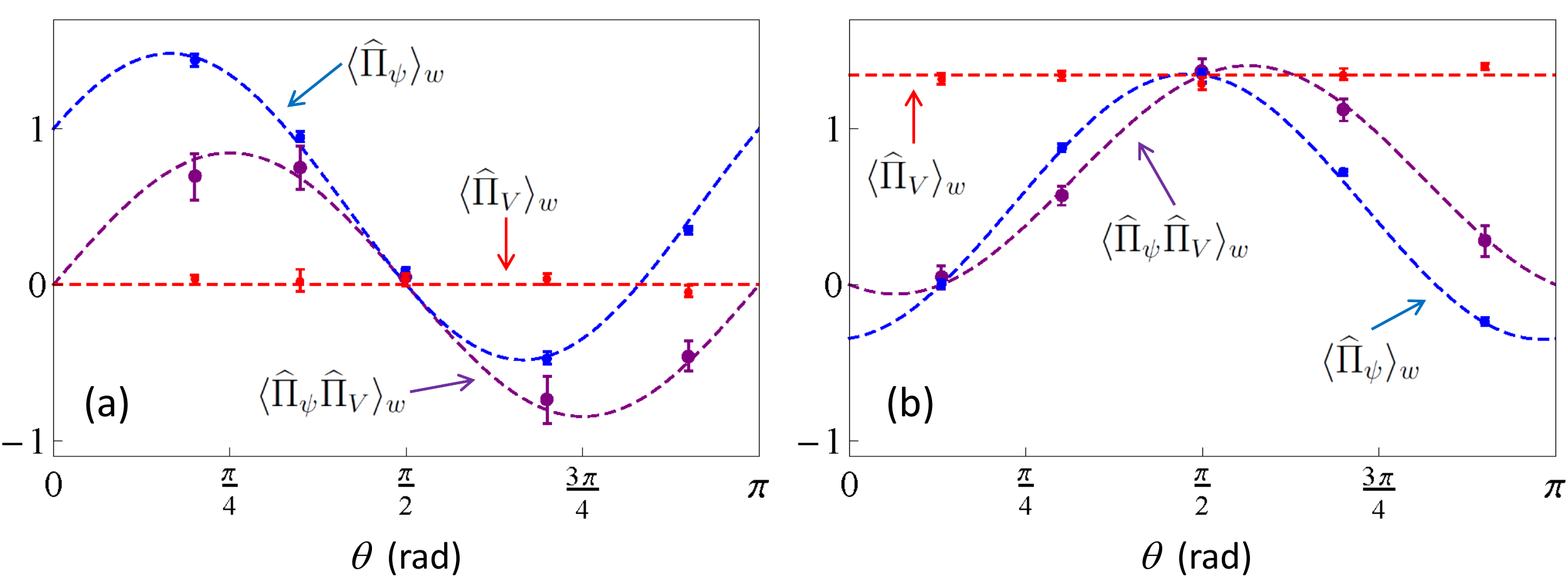}
\caption{Measured weak values (data points) compared with the theoretical predictions (dashed lines) for different $\widehat{\Pi}_\psi$ (i.e. for different values of $\theta$, since $|\psi\rangle = \cos \theta |H\rangle + \sin \theta |V \rangle $).
Blue and red points and lines correspond to the evaluations of the single-weak-value $\langle \widehat{\Pi}_\psi \rangle_w$ and  $\langle \widehat{\Pi}_V \rangle_w$, respectively, while purple points and line represent the evaluation of the sequential-weak-value $\langle \widehat{\Pi}_\psi \widehat{\Pi}_V   \rangle_w$.
Uncertainty bars are evaluated on the basis of sequences of repeated measurements. The uncertainty bars are naturally bigger in the case of the evaluation of sequential-weak-values with respect to the case single-weak-values, since in the former case the quantity measured is a covariance of positions, while in the latter cases they are position mean values.
The pre-selected and post-selected states are respectively $|\psi_i \rangle = 0.588 |H\rangle + 0.809 |V \rangle$ and $|\psi_f \rangle = |H\rangle$ for plot (a), and $|\psi_i \rangle = 0.509 |H\rangle + 0.861 |V \rangle $ and $|\psi_f \rangle = -0.397 |H\rangle + 0.918 |V \rangle$ for plot (b).
} \label{results} \end{center} \end{figure}
\end{widetext}

\end{document}